%% file: paper.tex
\documentclass[manuscript]{acmart}

\usepackage[colorinlistoftodos]{todonotes}
\usepackage{dsfont}
\usepackage{dirtytalk}
\usepackage{subfig}
\usepackage{svg}
\usepackage{multirow}

\newcommand{\STAB}[1]{\begin{tabular}{@{}c@{}}#1\end{tabular}}

\newcommand{\lexicon}{\mathcal{L}}

\def\bsq#1{%both single quotes
\lq{#1}\rq}

\AtBeginDocument{%
  \providecommand\BibTeX{{%
    \normalfont B\kern-0.5em{\scshape i\kern-0.25em b}\kern-0.8em\TeX}}}

\begin{document}

\title{The Usability of Pragmatic Communication in Regular Expression Synthesis}
%\title{Regular Expression Synthesis as Pragmatic Communication}
%\title{Regular Expression Synthesis as Pragmatic Communication}

% \author{Priyan Vaithilingam}
% \email{pvaithilingam@g.harvard.edu}
% \author{Elena Glassman}
% \email{eglassman@g.harvard.edu}
% \affiliation{%

% \affiliation{%

\author{Priyan Vaithilingam}
\email{pvaithilingam@g.harvard.edu}
\affiliation{%
  \institution{Harvard University}
  \city{Cambridge}
  \country{USA}
}

\author{Yewen Pu}
\email{yewen.pu@autodesk.com}
\affiliation{%
  \institution{Autodesk Research}
  \city{San Francisco}
  \country{USA}
}

\author{Elena Glassman}
\email{eglassman@g.harvard.edu}
\affiliation{%
  \institution{Harvard University}
  \city{Cambridge}
  \country{USA}
}

\renewcommand{\shortauthors}{Vaithilingam et al.}

\begin{abstract}

Programming-by-example (PBE) systems aim to alleviate the burden of programming. However, user-specified examples are often ambiguous, leaving multiple programs to satisfy the specification. Consequently, in most prior work, users have had to provide additional examples, particularly negative ones, to further constrain the search over compatible programs. Recent work resolves additional ambiguity by modeling program synthesis tasks as pragmatic communication, showing promising results on a graphics domain using a rudimentary user-study. We adapt pragmatic reasoning to a sub-domain of regular expressions and rigorously study its usability as a means of communication both with and without the ability to provide negative examples. Our user study (N=30) demonstrates that, with a pragmatic synthesizer, end-users can more successfully communicate a target regex using positive examples alone (95\%) compared to using a non-pragmatic synthesizer (51\%). Further, users can communicate more efficiently (57\% fewer examples) with a pragmatic synthesizer compared to a non-pragmatic one.

\end{abstract}

%    <concept>
%        <concept_id>10003120.10003121.10003129</concept_id>
%        <concept_desc>Human-centered computing~Interactive systems and tools</concept_desc>
%        <concept_significance>500</concept_significance>
%        </concept>
%  </ccs2012>

% \ccsdesc[500]{Human-centered computing~Interactive systems and tools}



\maketitle

\input{sec_introduction}

\input{sec_related}

\input{sec_background}

\input{sec_background_2}

\input{sec_implementation}

\input{sec_study}

\input{sec_discussion}

\begin{acks}
Statistical support was provided by data science specialist Noah Greifer, at the Institute for Quantitative Social Science, Harvard University.
\end{acks}

\bibliographystyle{ACM-Reference-Format}
\bibliography{references}

\newpage
\appendix
\input{sec_appendix}

\end{document}

%% file: sec_introduction.tex
\section{Introduction}

% As computers become more integrated into our daily lives, more end-users are exposed to tasks that are best solved via programming than ever before. For instance, one might want to go through a text file and extract all the valid addresses, which is typically solved by writing regular expressions (regexes henceforth). 

Regular expressions, or regexes, are a versatile text processing utility that has numerous applications ranging from search and replacement to input validation. Despite their widely recognized utility, regexes are considered hard to understand and compose, making even experienced users prone to errors~\cite{Chapman2017ExploringRE, Chapman2016ExploringRE, spishak2012typesystem, LouisGMichael2019RegexesAH}. 
%How might one alleviate the burden of programming, for whom the task of programming---such as writing regexes---is difficult? 

\emph{Programming by example (PBE)} can ease the challenge of getting to a desired program, e.g., a desired regex, because instead of writing the program from scratch, the user can provide examples of desired outputs for particular inputs~\cite{gulwani2015inductive} and a PBE system may return one or more partial~\cite{sslpbe} or complete programs that return those desired outputs on some~\cite{peleg2020perfect} or all of those particular inputs. 
%are generated for the user based on a specification that the user provides, i.e., examples of desired outputs for particular inputs~\cite{gulwani2015inductive}. 
Communicating with the computer about a desired program's behavior by giving input-output examples is powerful because these examples are both relatively easily understood by end-users and precisely verifiable by machines. Consequently, many state of the art program synthesizers that generate complex programs follow the PBE paradigm~\cite{solar2006combinatorial,ellis2019write,gulwani2011automating}.

A key challenge of using examples as a form of specification is that examples given by end-users are often ambiguous. Given the set of examples, there might be many different programs that simultaneously satisfy the specification. Resolving this ambiguity is crucial for a PBE system to be successful. Without resolving ambiguity, the synthesizer will generate spurious programs that fail to behave as the user desires on additional inputs. To address this challenge, some PBE systems may require users to provide additional examples to resolve these ambiguities~\cite{solar2006combinatorial, polozov2015flashmeta, le2017interactive}. However, prior work has shown that users are reluctant to provide more than a few examples~\cite{lau2009programming}. Some works have proposed interactive approaches that enable users to directly refine synthesized programs rather than crafting additional examples~\cite{mayer2015user, kandel2011wrangler, hila2018}, but editing regular expressions has been shown to be as difficult as writing regular expressions~\cite{LouisGMichael2019RegexesAH}. 

Some prior work has introduced inductive biases to resolve ambiguities, such as preferring \textit{smaller programs}~\cite{alur2013syntax}, or using a domain specific ranking functions~\cite{gulwani2011automating}, but these inductive biases may or may not be mentally modelable to end-users. If they are unintuitive, they could contribute to a mismatch between the user's implicit expectations and the capabilities of the synthesizer---which prior work has termed the \emph{user-synthesizer gap}~\cite{sslpbe}.  

A recently proposed inductive bias~\cite{pu2020program} based on pragmatic communication between humans~\cite{frank2012predicting} may be more objectively and subjectively usable, i.e., less likely to cause significant user-synthesizer gaps. Here the program synthesis task is modeled as pragmatic communication between the users and the synthesizer. During the synthesis process, when faced with multiple programs that satisfy the specification, the synthesizer reasons, via a mathematical model of pragmatic communication, about the question \say{\emph{Why} would a pragmatic speaker (the user) use this particular specification to communicate that program?} In doing so, the pragmatic synthesizer may select the program that is more likely to be the one the user intends to communicate with the given specification. \citet{pu2020program} demonstrated that, over an artificial graphical domain, users were able to communicate the intended programs to a pragmatic synthesizer with fewer examples relative to a non-pragmatic synthesizer. 

However, the full experience of interacting with pragmatic synthesis was not captured in enough detail by ~\citet{pu2020program} to sufficiently understand the usability of this approach in detail, i.e., the advantages and disadvantages from a user's perspective. Given how computationally expensive the approach is without further algorithmic innovations, it is necessary to study the user experience in sufficiently satisfying detail before any continued investment in this technique is made by members of the HCI community.

% However, they only evaluate pragmatic synthesis over a single domain and it is an open question as to whether pragmatic synthesis will be usable in other domains.

We instantiate pragmatic PBE synthesis in a subdomain of regular expressions for the first time, in order to characterize the user experience of PBE with a pragmatic inductive bias in a domain where PBE has been helpful in the past~\cite{bartoli2014,claver2021regis,li2021transregex,ye2020sketch,zhang20,zhang21}. However, we restricted the synthesizer to a sufficiently small subdomain of regular expressions to make the computation currently necessary for the pragmatic approach practical and interactive. This limited sub-domain of regular expressions only supports two literals ($0,1$) and three operators ($*$, $+$, ${n}$). 

Within this subdomain of regular expressions, we ran a controlled user study ($N=30$) on the objective and subjective experience of pragmatic PBE users. We also broke apart that usability as a function of whether or not users could give only positive or both positive and negative examples. %Unlike the first work on pragmatic synthesis (\citet{pu2020program}), we instantiate pragmatic synthesis with and without the affordance for users giving both positive and negatives examples, and in a new domain, i.e., regular expressions. 
%\todo{write a topic sentence for this paragraph}
We found that 86.6\% of participants preferred pragmatic synthesis over non-pragmatic synthesis. This is perhaps because they completed their assigned communication tasks with 44\% fewer examples ($\sim$2.5 examples less) when communicating with the pragmatic synthesizer.
We hope that strong positive usability characteristics revealed through this study will motivate future investment in developing the scalability of pragmatic synthesis.%, as its current implementation required us to tightly restrict the domain of regular expressions in this study. %Our restricted regular expressions domain only includes simple components, and characters supported (discussed in section \ref{sec:regdomain}).

%In this work, we formally . . . 
% we adapt Pu et al.'s pragmatic program synthesis to a new domain, specifically 
%on a sub-domain of regular expressions, and study its usability as a means of communication with the synthesizer.
%[overall say what we do here instead comparing to pu's prior work too much ]. With a user study ($N=30$), we replicate the prior finding that users can communicate programs with fewer examples to a pragmatic synthesizer than to a non-pragmatic synthesizer, and further characterize the differences in user experience and performance while communicating with both synthesizer types---with and without the ability to provide negative examples. 
%We also show that users can successfully communicate to the pragmatic synthesizer using positive examples alone.  

In summary, we make the following contributions:

\begin{itemize}
    \item We adapted and implemented pragmatic synthesis for a small sub-domain of regular expressions, supporting both positive and negative examples. 
    \item We examined the objective and subjective usability of pragmatic synthesis in a controlled user study, directly comparing pragmatic and non-pragmatic synthesis, with and without positive and negative examples.
    \item We found that, while participants must resort to generating negative examples for the non-pragmatic synthesizer, which they found cognitively challenging, participants were able to communicate their intent to the pragmatic synthesizer just as well with only positive examples. %\todo{we can make our results sound stronger}
\end{itemize}

%% file: sec_related.tex
\section{Related Work}

%We cover PBE, regex, pragmatics.

\subsection{Programming By Example}
Programming by example (PBE) was primarily developed in the programming languages and formal methods communities, e.g.,~\cite{solar2006combinatorial,gulwani2011automating,gulwani12,singh2015predicting,polozov2015flashmeta,leung15,feser2015synthesizing}, and occasionally in the AI and ML communities, e.g.,~\cite{lau2003programming,pu2020program,ellis2019write}. It has also been studied extensively in HCI contexts. Some of the earliest work in the HCI community describing or using PBE and exploring both its usability challenges and opportunities include ~\citet{myers86},~\citet{myers91},~\citet{cypher91},~\citet{fujishima98},~\citet{wolber02},and~\citet{cacm2000}. More recently, work has been published in the HCI community on the usability of modern PBE systems, e.g.,~\cite{le2017interactive,lau2009programming}, as well as novel interfaces and interaction models to better support PBE users, e.g.,~\cite{yessenov13,mayer2015user, kandel2011wrangler, hila2018,sslpbe,zhang20,zhang21}. This work extends this prior work in HCI by exploring pragmatic PBE's usability and the user experience of different affordances.

A long-standing problem in PBE is the inherent ambiguity in user-provided examples. Most PBE systems require the users to provide more examples to reduce the ambiguity. Some prior work in programming languages and machine learning communities use hand-crafted or machine-learned inductive biases to resolve the ambiguities. These inductive biases can be hand-crafted heuristics or ranking functions \cite{gulwani2011automating}, distance-based objective functions \cite{qlose2016} or probabilistic models learned from a large corpora \cite{raychev2016learning, DBLP:journals/corr/abs-1906-10816}. Pu et al. \cite{pu2020program} recently introduced an inductive bias based on pragmatic reasoning in an artificial graphical program domain. We build on this work by translating it to a new domain, and by adding the ability for users to provide positive and negative examples. %Our work does not intend replace the inductive biases previously proposed, but rather augment it.

%\cite{Santolucito19} 
%\cite{narita19} 
%\cite{wang17} 

%~\cite{alur2013syntax} 

\subsection{Pragmatics}
The ability for humans to resolve ambiguous utterances in a given context is called \emph{pragmatics}, first formally studied by cognitive linguists.%~\citep{grice1975logic}.
Initially, rational communication between humans was modeled by a set of heuristic-like maxims such as truthfulness, parsimony, and relevancy~\cite{grice1975logic}. 

Rational Speech Acts (RSA)~\cite{frank2012predicting} is a more modern model of pragmatic communication between humans. The RSA algorithm provides a computational model of pragmatic behaviours exhibited by human speakers and listeners by formalizing pragmatics in terms of Bayesian reasoning. The RSA algorithm has already been adopted in the field of natural language processing, where language descriptions were used to pragmatically infer a target referent, either from a set of categorical outcomes~\cite{monroe2017colors, andreas2016reasoning} or from a small set of programs~\cite{wang2016learning}. \citet{pu2020program} applies an incremental pragmatic program synthesis algorithm to a grid layout domain, using input-output examples as the means of communication between end-users and the program synthesizer, demonstrating that pragmatic synthesis is possible. Our work is the first work studying the usability of pragmatic reasoning for program synthesis from an HCI standpoint.%, with extensive user studies and analysis in a sub-domain of regular expression synthesis.

\subsection{Regular Expressions and Synthesis}

Regular expressions are a versatile text-processing utility with numerous applications for both programmers and end-users. Regular expressions are known to be hard to compose, understand, as well as error prone even for experienced users~\cite{Chapman2017ExploringRE, Chapman2016ExploringRE, spishak2012typesystem, LouisGMichael2019RegexesAH}. There have been a number of regular expression synthesizers in the past from both the programming language community and the machine learning community~\cite{bartoli2014, chen2020multi, ye2020sketch, claver2021regis}. %While prior works show the importance of having a usable regular expression synthesizer, they do not evaluate usability. 

There have been prior work in the HCI community to understand the usability of regex synthesizers and extend the interactive techniques originally proposed with the PBE systems. For instance,~\citet{zhang20} introduces augmented examples to resolve ambiguities. Interpretable program synthesis work by~\citet{zhang21} extends this interaction to include feedback from the synthesizer to help users successfully guide the synthesizer to the solution. 
By studying the usability of pragmatic synthesis in the domain of regular expressions, we show that the pragmatic synthesis framework can potentially complement these prior works by enabling the synthesizer to pragmatically reason about users' inputs.

%% file: sec_background.tex
\section{Background: Pragmatic Synthesis} \label{sec:pragsynth}

In this section, we cover \citet{pu2020program}'s approach to applying Rational Speech Acts to program synthesis, which we use to instantiate pragmatic synthesis in a subdomain of regular expressions. 

\subsection{Rational Speech Acts (RSA)}
\begin{figure}[h]
  \centering
  \includegraphics[width=\linewidth]{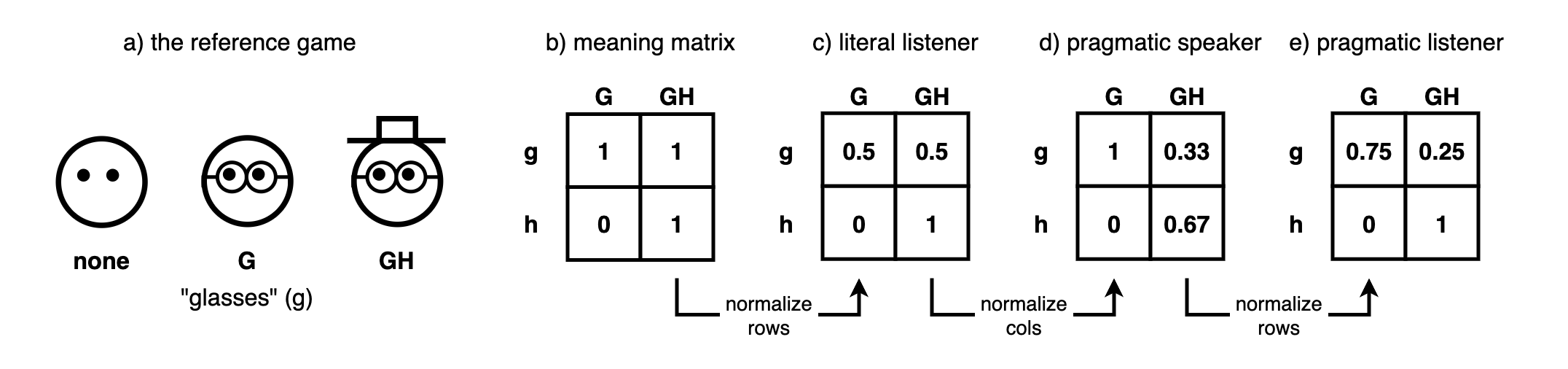}
  \caption{A simple example of pragmatic communication using rational speech acts}
  \label{fig:simple_prag}
\end{figure}

%The algorithm behind pragmatic inference, Rational Speech Acts~(RSA)~\cite{frank2012predicting, goodman2016pragmatic}, serves as the backbone of our pragmatic regex synthesizer. 
Rational Speech Acts (RSA) mathematically models how humans use contextual observations or knowledge to resolve ambiguity.
%\subsection{An Example}
Consider two human conversation partners, Alex and Basel, who are both looking at three cartoon faces, shown in  Figure~\ref{fig:simple_prag}(a): the first cartoon is wearing nothing (none), the second cartoon is wearing glasses (G), and the third cartoon is wearing both glasses and a hat (GH). In this (constrained) conversation, Alex or Basel can verbally point to which cartoon they are referring to by using one of the following descriptions: ``none (n)'', ``glasses (g)'', or ``hat (h)''. If Alex says ``glasses (g)'', RSA predicts that the human listener, Basel, would perform pragmatic inference and guess that Alex is referring to the cartoon G that wears \emph{only} glasses, rather than the cartoon GH that wears both glasses \emph{and} a hat. The intuition behind this is that if GH was the cartoon being referred to, Alex would have said ``hat (h)'' instead. 

%\subsection{Reference Game}
RSA models this pragmatic inference as recursive Bayesian reasoning, under the framework of a \emph{reference game}.
In a reference game, a speaker-listener pair $(S,L)$ cooperatively communicate a concept $w \in W$ using utterances $u \in U$ over a shared meaning matrix $\lexicon$, which denotes the consistency relationship between concepts and utterances: $\mathds{1}(w \vdash u)$. In the previous example, Alex played the role of a speaker, while Basel played the role of a listener. The space of concepts is $\{none, G, GH\}$. The space of utterances is $\{n, g,h\}$. The meaning matrix $\lexicon$ is shown in Figure~\ref{fig:simple_prag}(b), indicating if a hypothesis-utterance pair is consistent. For instance,  $\mathds{1}(G \vdash g) = 1$ (hypothesis ``G'' is consistent with the utterance ``g''), and $\mathds{1}(G \vdash h) = 0$ (hypothesis ``G'' is not consistent with the utterance ``h''). Given a concept $w$, the speaker $S$ chooses an utterance $u$ to communicate the concept. The communication is successful if the original concept that the speaker was trying to refer to is correctly identified by the listener.

\subsubsection{Recursive Bayesian Reasoning}

The RSA algorithm constructs a pragmatic listener that approximates the human's communicative behaviours by building an alternate chain of listener and speakers. As illustrated in Figure~\ref{fig:simple_prag}(b-e), starting from the meaning matrix $\lexicon$, RSA first constructs a probability distribution of the literal listener $L_0$, then that of a pragmatic speaker $S_1$, and finally a pragmatic listener $L_1$, where each distribution is recursively defined over the previous one. 

\noindent
\textbf{Literal listener $L_0$}: Given an utterance $u$, the literal speaker defines a conditional distribution of concepts as follows: $L_0(w|u) \propto P(w) \mathds{1}(w \vdash u)$. That is to say, the probability of selecting a concept $w$ given utterance $u$ is proportional to its prior distribution $P(w)$, and whether the concept-utterance pair is consistent $\mathds{1}(w \vdash u)$. If we assume the prior is uniform, i.e., having no global preference over one concept over another, the proportionality amounts to marginalizing across all alternative concepts $w'$: $L_0(w|u) = \frac{\mathds{1}(w \vdash u)}{\sum_{w'} \mathds{1}(w' \vdash u)}$, which puts a uniform distribution over all the consistent concepts. Figure~\ref{fig:simple_prag}(c) shows the literal listener for example reference game, obtained by marginalizing each \emph{row} of the meaning matrix $\lexicon$, and where each row denotes the conditional distribution $L_0(w|u)$. This listener, upon hearing the utterance $g$, will select either hypothesis $G$ or $GH$ with equal probability, which does not capture pragmatic behaviour of favoring $G$ over $GH$.

\noindent
\textbf{Pragmatic speaker $S_1$}: Building on top of the literal listener $L_0$, RSA recursively constructs a pragmatic speaker $S_1$ as follows: $S_1(u|w) \propto L_0(w|u)$. That is to say, given a concept $w$, the pragmatic speaker $S_1$ will select utterance $u$ proportional to how likely the literal listener, $L_0$, is going to recover the concept $w$ upon receiving $u$. This amounts to marginalizing each \emph{column} of $L_0$. The pragmatic speaker $S_1$ is given in Figure~\ref{fig:simple_prag}(d) where each column denotes the conditional distribution $S_1(u|w)$. As we can see, given the hypothesis $GH$, the pragmatic speaker is more likely to select the utterance $h$ over $g$ as the literal listener $L_0$ is more likely to recover $GH$ upon receiving $h$. 

\noindent
\textbf{Pragmatic listener $L_1$}: RSA recursively constructs the pragmatic listener $L_1$ as follows: $L_1(w|u) \propto S_1(u|w)$. That is to say, given an utterance $u$, the pragmatic listener $L_1$ is more likely to select a concept $w$ if the pragmatic speaker $S_1$ is more likely to use $u$ to communicate $w$. This amounts to normalizing the \emph{rows} of $S_1$. Figure~\ref{fig:simple_prag}(e) shows how the pragmatic listener $L_1$ has the desired behaviour of preferring hypothesis $G$ over $GH$ upon receiving the utterance $g$. While it is possible to continue this recursive construction, prior work~\cite{degen2013cost} suggests that humans do not continue this recursive construction as it is computationally costly.

%% file: sec_background_2.tex
\subsection{Programming By Example (PBE) With Pragmatics}

\citet{pu2020program} cast Programming By Example as an instance of rational communication by drawing a series of analogies between concepts of communication and that of program synthesis. In short, programming by examples is a communicative act where end-users ($S$) attempt to communicate a target program ($w$) among the space of all possible programs ($W$) to the synthesizer ($L$) by giving the synthesizer a sequence of examples ($D$). Table~\ref{tab:analogy} lays out these analogies. 

The synthesizer is implemented as a pragmatic listener ($L_{1}$) which calculates the probability $P_{L_1}(w|D) \ \forall w \in W$, where $W$ is the list of all the concepts. We choose $argmax_{w}(P_{L_1}(w|D))$ as the synthesized regex (see \ref{ap:algorithm}). If there are multiple concepts that have the same maximum probability, then we choose one of them at random.

\begin{table}[h]
\centering
\begin{tabular}{|l|l|l|l|l|l|l|l|}
\hline
                                                                  & $S$        & $L$           & $W$                                                           & $D$          & $\lexicon$              & $L_{0}$                                                               & $L_{1}$                                                               \\ \hline
\textbf{\begin{tabular}[c]{@{}l@{}}rational \\ communication\end{tabular}} & speaker  & listener    & concepts                                                    & utterances & meaning matrix & \begin{tabular}[c]{@{}l@{}}literal \\ listener\end{tabular}      & \begin{tabular}[c]{@{}l@{}}pragmatic \\ listener\end{tabular}    \\ \hline
\textbf{\begin{tabular}[c]{@{}l@{}}programming \\ by example\end{tabular}} & end-user & synthesizer & \begin{tabular}[c]{@{}l@{}}programs \\ (regex)\end{tabular} & examples   & satisfiability & \begin{tabular}[c]{@{}l@{}}SAT-based \\ synthesizer\end{tabular} & \begin{tabular}[c]{@{}l@{}}pragmatic \\ synthesizer\end{tabular} \\ \hline
\end{tabular}
\caption{A list of analogies between rational communication and programming by examples}
\label{tab:analogy}
\end{table}

Under this framing, one can view traditional SAT-based synthesizer \cite{solar2008program,lau2003programming,feser2015synthesizing} as a literal listener, which returns \emph{any} satisfying program to the end-user with equal probability. Approaches such as~\cite{polozov2015flashmeta,learningToRank,singh2015predicting} attempt to either (1) \emph{rank} the satisfying programs by a hand-crafted heuristic or (2) learn from historical usage data. 

In contrast,~\citet{pu2020program} applies the RSA algorithm to the meaing matrix of PBE to \emph{induce} a ranking, without having to manually define a heuristic nor having to learn from historical usage data. Formally, this approach constructs a pragmatic program synthesizer $L_1(w|D)$ which, upon receiving a \emph{set} of user-given examples $D$, computes a distribution over the space of all programs $W$ using the RSA algorithm. %\todo{clarify that order of examples in this sequence does not change robot guess} 
A key distinction between the vanilla RSA algorithm and this RSA algorithm for PBE is that in the PBE setting, the user can give \emph{multiple} examples instead of a single utterance. 
The approach outlined by~\citet{pu2020program} and~\citet{cohn2018incremental} models an \emph{incremental} pragmatic speaker that generates one example at a time to handle this sequence of user-given examples.

\section{Pragmatic Synthesis for a Subdomain of Regular Expressions}
We replicate the algorithmic approach of incremental pragmatic program synthesis of \citet{pu2020program} in a subdomain of regular expressions.

\subsection{Defining the Regular Expression Sub-Domain} \label{sec:regdomain}
As a system designer, to instantiate pragmatic PBE in any domain with the state of the art method described above, one must define the set of possible programs the speaker could possibly be referring to and the set of possible utterances they could provide to indicate it. These defined sets are finite, even though the space of possible utterances and programs that the speaker is referring to may not be, so pragmatic PBE will be constrained by the size of these sets, i.e., the number of rows and columns in the meaning matrix $\lexicon$. The larger these two sets, the more programs are reference-able by the speaker, but the more computationally demanding the calculation, because
%One limitation of performing pragmatic inference is that 
the pragmatic inference step requires computation over the entire cross-product space of concepts (all programs in $W$) and all utterances (all examples in $D$), making it intractable for large domains without further research into identifying and implementing suitable optimizations, exact or approximate.

Therefore, when choosing the set of concepts (regexes) and utterances (example strings), we chose as small set of each to make sure the system was sufficiently responsive in an interactive setting,
%We chose a limited set of regexes and examples to keep the algorithm interactive and computationally practical, 
while still retaining the core complexities associated with synthesizing regex such as ambiguity. In doing so, we could focus on the usability of pragmatic reasoning in the PBE domain of regexes, and leave the scalability of pragmatic reasoning as future work.
%\todo{elena continue here}

Specifically, we composed a set of concepts by defining a 
%\subsubsection{Set of regular expressions (Hypotheses/Concepts)}
%We considered a sub-domain of regular expressions generated from 
grammar containing two characters \bsq{0}, \bsq{1}, and four operations \bsq{*}, \bsq{+}, \bsq{\{1\}}, and \bsq{\{2\}}, shown in full in Table~\ref{tab:grammar}. Using this grammar, we generated 10000 regexes and randomly chose a subset of 350 to designate as the set of concepts over which pragmatic inference would be computed---the columns of the meaning matrix $\lexicon$. The set of regexes in this set varies from simple, e.g., \bsq{\textit{0\{2\}}}, to complicated, e.g., \bsq{\textit{[01]*[01]\{2\}[01]\{2\}}}.
%\subsubsection{Set of examples strings (Utterances)}
For the set of utterances the user could use to refer to any one of these 350 possible regexes, we enumerated 
%To make it easy for users to intuitively give examples, we choose 
all possible strings containing combinations of \lq$0$\rq and \lq$1$\rq up to 10 digits long. This defines a set of $2^{11}-1$ strings, which are the rows of the meaning matrix $\lexicon$.

\begin{table*}[]
\centering
\begin{tabular}{l}
Regular expression grammar \\ \hline
$S \rightarrow RP \ | \ S \ RP$ \\
$RP \rightarrow CHAR\ OP$\\
$OP \rightarrow \text{\bsq{*}}\ |\ \text{\bsq{+}}\ |\ \text{\bsq{\{1\}}}\ |\ \text{\bsq{\{2\}}}$\\
$CHAR \rightarrow \text{\bsq{[01]}}\ |\ \text{\bsq{0}}\ |\ \text{\bsq{1}}$\\
\end{tabular}
\caption{A regular expression grammar used to generate all the concepts (list of regexes) used in the user study. $CHAR$ represents valid characters, $OP$ represents regex quantifiers, $RP$ represents an atomic part of the regular expression. $S$ is the starting non-terminal which generates a valid regular expression.}
\label{tab:grammar}
\end{table*}

% We are not addressing the scalability of pragmatic reasoning, we leave that for future work

\subsection{Pragmatic PBE Example with Regular Expressions}
To illustrate how we operationalized RSA-based pragmatics 
%using the algorithm outlined in~\citet{pu2020program} and~\citet{cohn2018incremental} 
for regular expressions, we can first walk through a small example reference game: % where by-hand calculation is feasible:
% \subsection{Program Synthesis as a Reference Game}
% first make the analogy of utterances as examples, and hypothesis as programs. introduce incremental pragmatics, show you need that to handle multiple examples, show on a concrete example
Consider a reference game with four different concepts (regular expressions) $W = \{w_0, w_1, w_2, w_3\}$ and four utterances (example strings) $U = \{u_0, u_1, u_2, u_3\}$, shown in Table~\ref{tab:mm}. Each concept is a particular regular expression which works on characters \lq$0$\rq and \lq$1$\rq, and each utterance is an example string containing characters \lq$0$\rq and \lq$1$\rq. Table~\ref{tab:mm} is the meaning matrix $\lexicon$, where each entry $(i, j)$ denotes whether $w_j$ is consistent with $u_i$ ($w_j \vdash u_i$). %Given a set of examples $D$, we say $w \vdash D$ if $\forall u \in D, w \vdash u$.

In Table~\ref{tab:mm}, if the user wants to refer to the regex $w_0$ that requires the string to end with $0$ the can provide the set of example strings $D = \{u_1, u_2\}$. RSA. from the given set of examples, predicts that the person is trying to refer to $w_0$: 
Of the regular expressions making up the list of concepts in this meaning matrix, only $w_0$ and $w_3$ are consistent with $D$, but $w_0$ is the only regular expression that imposes a strong requirement for the string to end with the character $0$. However, regular synthesizer which only checks for consistency can either predict $w_0$ or $w_3$, even though $w_3$ is more general.

\textit{Synthesizer with support for positive and negative examples:}~
Pu et al.'s \cite{pu2020program} work on pragmatic synthesizer focuses on positive examples. In this work we also want to study the communication dynamics with negative examples -- examples that are inconsistent with the program. To achieve this, we modify the meaning matrix (see Table~\ref{tab:mm}) by changing the utterance from just example string to a pair of example string and a sign (positive vs negative examples). The new utterance is $(u_i, +/-)$. Then meaning matrix values at $(i, j)$ will be $w_j \vdash (u_i, +/-)$. An example of the modified meaning matrix is available in table \ref{tab:mmneg}. We then run the exact same pragmatic and non-pragmatic algorithms over the modified meaning matrix. In the UI, we tweak the input filed to append the sign (+/-) to user-provided examples.

\begin{table}
\centering
\begin{tabular}{cc|cccc}
                &               & \textbf{$w_0$}          & \textbf{$w_1$}     & \textbf{$w_2$}     & \textbf{$w_3$}        \\
                &               & \textbf{{[}01{]}+0+} & \textbf{1*0+1*} & \textbf{0*1+0*} & \textbf{{[}01{]}*} \\ \hline
\textbf{$u_0$}  & \textbf{1100} & 1                    & 1               & 1               & 1                  \\
\textbf{$u_1$}  & \textbf{0000} & 1                    & 1               & 0               & 1                  \\
\textbf{$u_2$}  & \textbf{0010} & 1                    & 0               & 1               & 1                  \\
\textbf{$u_3$}  & \textbf{0111} & 0                    & 1               & 1               & 1                 
\end{tabular}
\caption{This table shows the meaning matrix for a simple regular expression reference game. The concepts (regular expressions) are mentioned in column headers, and utterances (examples strings) are mentioned in row headers. Each cell $(i, j)$ denotes if example $u_i$ satisfies the regular expression $w_j$.}
\label{tab:mm}
\end{table}

\begin{table}
\centering
\begin{tabular}{cc|cccc}
                &               & \textbf{$w_0$}          & \textbf{$w_1$}     & \textbf{$w_2$}     & \textbf{$w_3$}        \\
                &               & \textbf{{[}01{]}+0+} & \textbf{1*0+1*} & \textbf{0*1+0*} & \textbf{{[}01{]}*} \\ \hline
\textbf{$u_1'$}  & \textbf{(1100, +)} & 1                    & 1               & 1               & 1                  \\
\textbf{$u_2'$}  & \textbf{(1100, -)} & 0                    & 0               & 0               & 0                  \\
\textbf{$u_3'$}  & \textbf{(0000, +)} & 1                    & 1               & 0               & 1                  \\
\textbf{$u_4'$}  & \textbf{(0000, -)} & 0                    & 0               & 1               & 0                  \\
\textbf{$u_5'$}  & \textbf{(0010, +)} & 1                    & 0               & 1               & 1                  \\
\textbf{$u_6'$}  & \textbf{(0010, -)} & 0                    & 1               & 0               & 0                  \\
\textbf{$u_7'$}  & \textbf{(0111, +)} & 0                    & 1               & 1               & 1                  \\
\textbf{$u_8'$}  & \textbf{(0111, -)} & 1                    & 0               & 0               & 0                 
\end{tabular}
\caption{This table shows the meaning matrix for a simple regular expression reference game that supports negative examples. The concepts (regular expressions) are mentioned in column headers, and utterances (pair of examples strings and its sign) are mentioned in row headers. Each cell $(i, j)$ denotes if example $u_i'$ satisfies the regular expression $w_j$.}
\label{tab:mmneg}
\end{table}

%% file: sec_implementation.tex
%\section{Design and Implementation}
%In this section we explain the design and implementation of the pragmatic synthesis algorithm and the interface used for the user study, using a restricted set of simple regexes and examples that permits real-time interactions.

%\subsection{Choosing a Sub-Domain} \label{sec:regdomain}

%In theory, we can scale this up to very large sets to make it virtually infinite. 
% However, recursive probabilistic reasoning is computationally very expensive, making it impractical. 

%% file: sec_study.tex
\section{User Study}

We conducted a user study to investigate the following research questions about users' objective success and subjective experience while using pragmatic synthesis, compared with non-pragmatic synthesis:

\begin{itemize}
    % \item \textbf{RQ1}: Do users communicate with the pragmatic synthesizer more effectively (number of examples) compared to the non-pragmatic synthesizer?
    %Does formulating regular expression synthesis as pragmatic communication help the user communicate with the synthesizer more effectively compared to traditional non-pragmatic synthesizer.
    \item \textbf{RQ1}: Which synthesizer (pragmatic or non-pragmatic) can guess the participant's intended regular expression from the fewest examples, when participants are limited to positive examples only? 
    \item \textbf{RQ2}: \textbf{RQ1}---when the interface allows participants to provide both positive and negative examples. %When participants can give both positive and negative examples, Can participants communicate a regex to the pragmatic synthesizer using fewer examples compared to traditional synthesizer, using both positive and negative examples?
    \item \textbf{RQ3}: How does the pragmatic inductive bias affect users' subjective experience?
    %How intuitive is it for the users to provide positive examples?
    %\item \textbf{RQ5}: How did the introduction of negative examples within pragmatic synthesis affect the user experience? %How does the performance of pragmatic synthesizer vary with a combination positive and negative examples compared to just positive examples?
\end{itemize}

\subsection{Participants}
We recruited 30 participants (10 female and 20 male) through several mailing lists of an R1 research university (25 participants), and through an announcement on Twitter (5 participants). Participants received a \$20 gift certificate for their time. The participant pool included one undergraduate student, nine Master's students, 17 Ph.D. students, and three participants from other professions. Twelve participants were between 18-25 years old, 17 participants were between 25-30 years old and one participant did not provide their age.
%\textit{Participant Experience}: 
Five participants reported they had more than five years of experience using regular expressions, eight reported 2-5 years of experience, and 17 reported less than a year of experience using regular expressions.
% \todo{do we have data on their ages?}
\begin{table}[]
\centering
\begin{tabular}{ccc}
\multicolumn{1}{l}{}                                   & \multicolumn{2}{c}{within-subjects}                                                                                                                                       \\ \cline{2-3} 
\multicolumn{1}{c|}{\multirow{2}{*}{\STAB{\rotatebox[origin=c]{90}{\begin{tabular}[c]{@{}c@{}}between- \\ subjects\end{tabular}}}}} & \multicolumn{1}{c|}{\begin{tabular}[c]{@{}c@{}}Non-pragmatic \\ UI(+)\end{tabular}}   & \multicolumn{1}{c|}{\begin{tabular}[c]{@{}c@{}}Pragmatic\\ UI(+)\end{tabular}}    \\ \cline{2-3} 
\multicolumn{1}{c|}{}                                  & \multicolumn{1}{c|}{\begin{tabular}[c]{@{}c@{}}Non-pragmatic\\ UI (+/-)\end{tabular}} & \multicolumn{1}{c|}{\begin{tabular}[c]{@{}c@{}}Pragmatic\\ UI (+/-)\end{tabular}} \\ \cline{2-3} 
\end{tabular}
\caption{This table shows the four different conditions participants performed their communication tasks.}
\label{tab:conditions}
\end{table}

\subsection{Conditions}

There are two independent factors in this experiment (see Table \ref{tab:conditions}). They are (1) the types of examples the participant can provide:
\begin{itemize}
    \item \textbf{$\mathbf{UI_{+}}$ (only positive examples)}: In this condition, the participants are only allowed to provide positive examples, i.e., strings that satisfy the target regular expression. 
    \item \textbf{$\mathbf{UI_{+/-}}$ (both positive and negative examples)}: In this condition, the participants are allowed to provide both positive examples, i.e., strings that satisfy the target regular expression, and negative examples, i.e., strings that do not satisfy the target regular expression, to the synthesizers.
\end{itemize}
and (2) the type of synthesizer they are communicating with:
\begin{itemize}
    \item \textbf{Non-Pragmatic Synthesizer}: 
    There are two broad categories of program synthesizers, First category of synthesizers returns the first program it finds that satisfies the specification~\cite{alur2017scaling, perelman2014test, albarghouthi2013recursive}. Second kind explores the search space to find K consistent programs and use a ranking function to pick the best of the K programs~\cite{gulwani2011automating, verbruggen2021semantic, rolim2017learning}. The most commonly used ranking function in prior research is to return the shortest program of the top k solutions~\cite{pan2021can, gulwani2014program}. However, due to the structure of our DSL, we use cannot use shortest program as our ranking function for our baseline. The regular expression $[01]^*$ is smallest regular expression which is consistent with any set of positive examples, therefore our synthesizer will always pick $[01]^*$ as the solution for any given set of examples. To prevent this, we use a random ranking function, where we pick a random program from the K consistent programs as the solution.
    %For the baseline condition, the state of the art synthesizers treats the user-generated examples literally (i.e. find any program that satisfies them). Most of these approaches uses top-down enumerative synthesis [cite, http://dx.doi.org/10.1145/2737924.2737977] or SAT-based solvers [cite] as it doesn't make a difference \emph{which} program to return to the user, as long as they're correct. However, to aid the search, there is an inductive bias to prefer smaller regular expressions.\todo{fit the next sentence somewhere before} [1. any satisfying program, 2. any sat program but shorter]
    % However, due to the nature of our DSL, the smallest regular expression that can be generated by the enumerative synthesizer is $[01]^*$ which is consistent will any set of positive examples. Therefore an enumerative synthesizer will always guess the regular expression $[01]^*$ as the solution for any given set of examples.
    Since we are only working with a finite set of regular expressions for this experiment, to accommodate our DSL, we implement a synthesizer based on the literal listener ($L_0$) model by using the same meaning matrix $\lexicon$. Let's say the list of examples provided by the user is $D$. The synthesizer is implemented as a literal listener ($L_{0}$) which calculates the probability $P_{L_0}(w|D) \ \forall w \in W$, where $W$ is the list of all the concepts. We choose $argmax_{w}(P_{L_0}(w|D))$ as the synthesized regex (see Appendix~\ref{ap:algorithm}). If there are multiple concepts that have the same maximum probability (i.e. there are multiple consistent regular expressions), then we choose one of them at random.
    
    \item \textbf{Pragmatic Synthesizer}: For the experiment condition, we implement a synthesizer based on the pragmatic listener ($L_1$) model by using the same meaning matrix $\lexicon$. Let's say the list of examples provided by the user is $D$.  The synthesizer is implemented as a pragmatic listener ($L_{1}$) which calculates the probability $P_{L_1}(w|D) \ \forall w \in W$, where $W$ is the list of all the concepts. We choose $argmax_{w}(P_{L_1}(w|D))$ as the synthesized regex (see Appendix~\ref{ap:algorithm}). If there are multiple concepts that have the same maximum probability, then we choose one of them at random.
\end{itemize}

\subsection{User Interface for the User Study}

\begin{figure}
    \centering
    \includegraphics[width=1\linewidth]{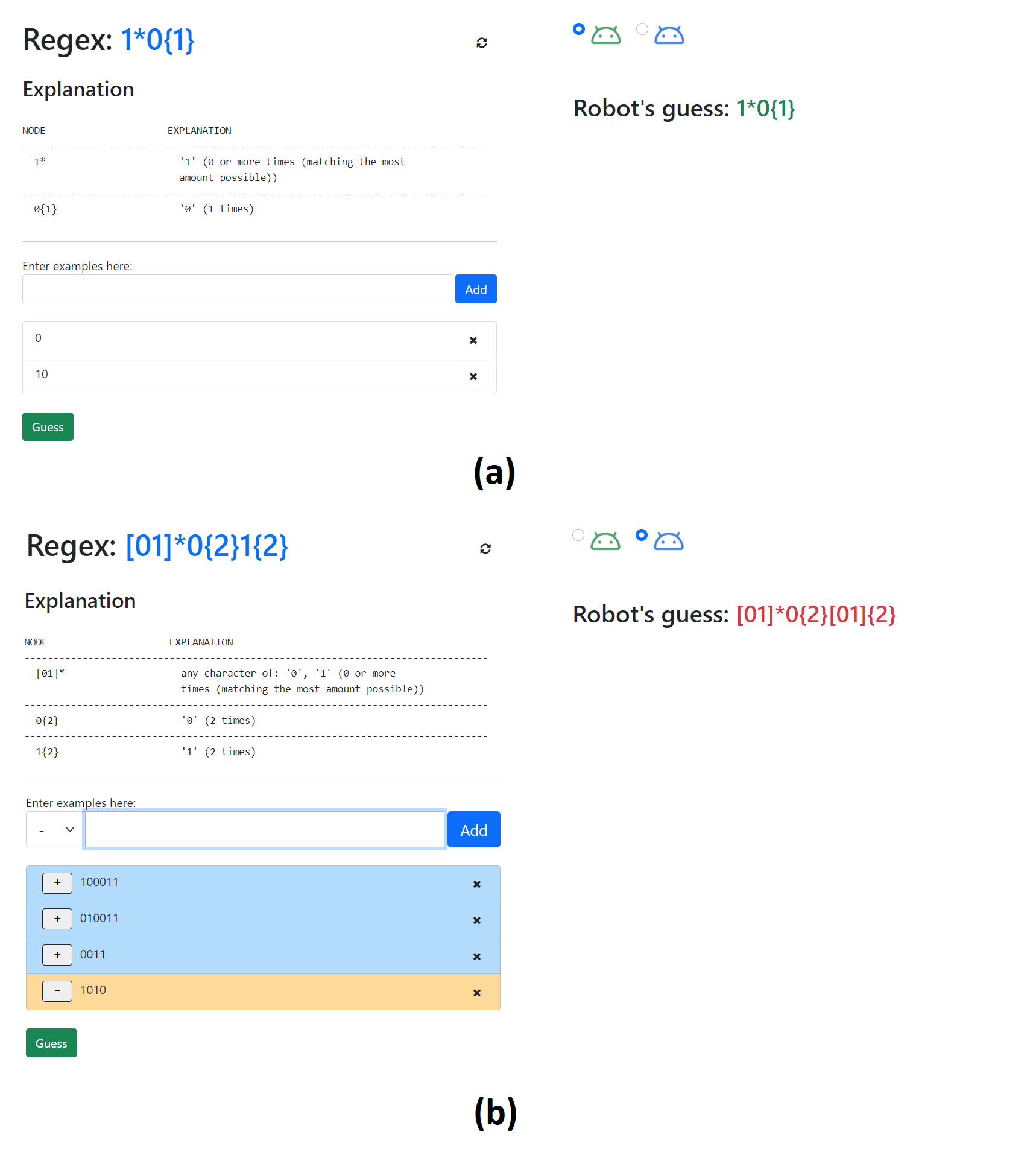}
    \caption{The two different user interfaces used in the study: (a) $UI_{+}$ and (b) $UI_{+/-}$. In both the interfaces, the target regular expression is shown in the top left corner in blue and a verbal description of the target regex is shown below. In the $UI_{+}$ interface (a), the users can provide only positive examples in the input field, whereas in the $UI_{+/-}$ interface (b), the user can add both positive and negative examples by using a drop-down menu. All the examples provided by the user are shown in the list below the input field, and the user can delete any example by pressing the \textbf{x} button.}
    \label{fig:ui}
\end{figure}

In order for participants to see and complete the communication task in any one of the assigned conditions, we implemented two web-based interfaces shown in Figure~\ref{fig:ui}, where the first interface only accepts positive examples, and the second interface accepts both positive and negative examples. %There are two example-giving conditions that the interface supports: (1) the interface only affords positive examples (Figure~\ref{fig:ui}(a)) and (2) the interface affords provide both positive and negative examples (Figure~\ref{fig:ui}(b)). 
Through their assigned interface, participants provide examples to communicate regular expressions to a pragmatic synthesizer and a non-pragmatic synthesizer, in a counterbalanced order. Which synthesizer the participant is communicating with is determined by which---blue or green---``robot'' they are instructed to choose during the course of their session. (Participants are blind to which robot is pragmatic and which is non-pragmatic.) Regardless of which robot they are communicating with, the robot's best guess is shown on the right hand side of the interface. 

Whenever the interface loads, it randomly picks a target regular expression from those in $W$ and displays it at the top of the left half of the interface. With this assigned target regular expression comes an automatically generated explanation of each of its components.

To communicate the assigned regex to the synthesizer "robot", participants can enter example strings of zeros and ones into the example entrance text-box. The set of the examples is initially empty. All the examples that are currently part of the set the participant is using to communicate their assigned regex to the robot are listed below the example entrance text-box. An "x" button allows participants to remove examples that they previously added.

The robot automatically updates its guess with every addition or removal of an example. Since the guess sometimes does not change and participants are sometimes not sure whether the robot did actually update its guess, an explicit ``Guess'' button is present for participants to optionally click, which increases their certainty that the robot has responded to their latest example set update.

\begin{figure}
    \centering
    \includegraphics[width=1.0\linewidth]{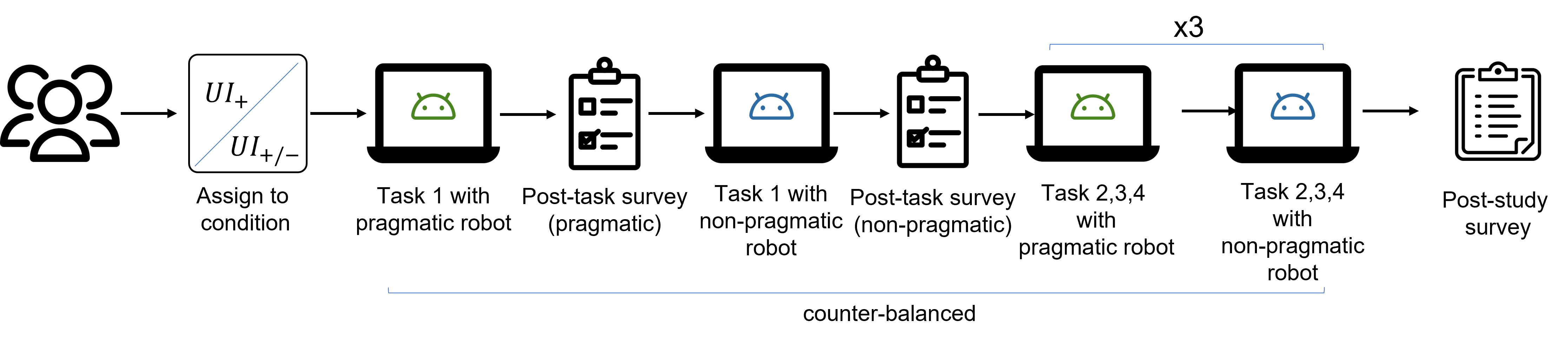}
    \caption{User study procedure. Both the UI conditions and robot order are counter-balanced. }
    \label{fig:procedure}
\end{figure}

\subsection{Procedure}  
To answer the above research questions, we conducted a mixed-design user study with 30 participants (see Figure~\ref{fig:procedure}). Each participant was assigned to one of the two UI conditions ($\mathbf{UI_{+}}$/$\mathbf{UI_{+/-}}$). The participant will use this UI for the rest of the user study, and the assignment of the UI conditions were counter-balanced. The participant is then given a tutorial on how to use the assigned UI for the study. Following the tutorial, the participant will start with the first communication task - where they have to communicate the target regular expression to the pragmatic robot. Once they completed or chose to stop, the participant filled a post-task survey describing their experience with the task. Then they proceeded to communicating the same target regular expression now to the non-pragmatic robot. Once they completed or choose to stopped, they will fill out the post-task survey for their experience with the non-pragmatic robot. The order of the robots are counter-balanced between the participants. This marks the completion of the first communication task. Following this, they will proceed to perform four more communication tasks with both pragmatic and non-pragmatic robot (the order of the robot is flipped with each task), however, the participants will not fill out anymore post-task surveys. After all the four communication tasks are completed, the participant filled out a post-study survey describing their overall experience with both the robots.

Throughout the study, participants were encouraged to think aloud. We recorded their screen, i.e., their interactions with the assigned user interfaces, along with the audio of the session. We later transcribed the recorded audio to be used in the qualitative analysis. The first-author performed open-coding on the audio transcripts and the participant's survey responses to identify themes which are used explain results in the qualitative result section.

\subsection{Measurements} %We measured the following quantitative and qualitative metrics during the user study.
We recorded both quantitative and qualitative metrics during the user study. To answer \textbf{RQ1} and \textbf{RQ2}, we recorded the \emph{number of examples} the participants provided in order to complete each task. We also recorded whether or not they were successful in completing each task. For the qualitative metrics, the first author open coded the participants' textual survey responses and relevant quotes from the transcribed audio.
%To capture the subjective experience of using both the pragmatic and non-pragmatic synthesizers, the participants fill out a post-task survey after communicating with the pragmatic robot and the non-pragmatic robot during the first communication task. 

Both the post-task and post-study surveys collect participants' subjective experiences, to answer \textbf{RQ3}.
The post-task survey questions (Table~\ref{tab:posttask}) include questions about their \emph{confidence in their chosen examples}, and the \emph{ease of communication} they experienced with the robot (synthesizer) they had just communicated with, and the NASA TLX questionnaire~\cite{hart1988development} on a 7-point Likert scale. 
The post-study survey questions (Table~\ref{tab:poststudy}), which participants filled out after communicating with both synthesizers multiple times, included user's self-reported \emph{preference} among the robots and modified NASA TLX questions that focused on directly comparing their experience between green and blue robots (pragmatic and non-pragmatic synthesizers). %We also measure user's self-reported \emph{preference} among the robots. The questions are available in Table~\ref{tab:poststudy}.

% \noindent \textit{Interface metrics}: We recorded the \emph{Number of examples} the participants provided for each task and the task \emph{success/failure} in both the conditions.

% \noindent \textit{Post-task survey}: After the first task, we measure the initial experience of the participant using relevant questions from the NASA TLX questionnaire \cite{hart1988development}, i.e., \emph{Mental demand, Hurry/Rush, Effort, Success, and Frustration}. We also measure the participant's self reported \emph{confidence in the examples} they provided, and the \emph{ease of communication} with the robot. The questions asked in the post-task survey are available in Table~\ref{tab:posttask}. This survey is only filled by the participant after the first task. 

% \noindent \textit{Post-study survey}: After the participant completes all the four tasks, we measure relevant questions from the NASA TLX questionnaire slightly modified to compare the two robots, still covering \emph{Mental demand, Hurry/Rush, Effort, Success, and Frustration}. We also measure user's self-reported \emph{preference} among the robots. The questions are available in Table~\ref{tab:poststudy}.

Since our user study has both within-subject component (pragmatic vs. non-pragmatic synthesizers) and a between subjects component ($\mathbf{UI_{+}}$ vs. $\mathbf{UI_{+/-}}$), we run repeated measures MANOVA~\cite{Tabachnick2011} to report on statistical significance.

\begin{table*}
\begin{tabular}{l} \hline
Q1.1. How confident were you with the examples you provided? (1-Not confident, 7 - Very confident)\\
Q1.2. How easy was it to communicate with the robot? (1-Very hard, 7-very easy)\\
Q1.3. How was your communication experience with the robot? (Short textual answer)\\
Q1.4. What strategies did you use to communicate with the robot? (Short textual answer)\\
Q2.1. How mentally demanding was this task with this tool? (1—Very Low, 7—Very High)\\
Q2.2. How hurried or rushed were you during this task? (1—Very Low, 7—Very High)\\
Q2.3. How successful would you rate yourself in accomplishing this task? (1—Perfect, 7—Failure)\\
Q2.4. How hard did you have to work to accomplish your level of performance? (1—Very Low, 7—Very High)\\
Q2.5. How insecure, discouraged, irritated, stressed, and annoyed were you? (1—Very Low, 7—Very High)\\\hline
\end{tabular}
\caption{After the first communication task, participants rated (or a 7-point Likert scale) their communication experience (questions 1.1 - 1.4) and the subjective workload using NASA TLX measures (questions 2.1 - 2.5) for each robot. Participants rated on a 7-point scale.\label{tab:posttask}}
\Description[Post-task survey questions]{Post-task survey questions:
Q1.1. How confident were you with the examples you provided? (1-Not confident, 7 - Very confident)\\
Q1.2. How easy was it to communicate with the robot? (1-Very hard, 7-very easy)\\
Q1.3. How was your communication experience with the robot? (Short textual answer)\\
Q1.4. What strategies did you use to communicate with the robot? (Short textual answer)\\
Q2.1. How mentally demanding was this task with this tool? (1—Very Low, 7—Very High)\\
Q2.2. How hurried or rushed were you during this task? (1—Very Low, 7—Very High)\\
Q2.3. How successful would you rate yourself in accomplishing this task? (1—Perfect, 7—Failure)\\
Q2.4. How hard did you have to work to accomplish your level of performance? (1—Very Low, 7—Very High)\\
Q2.5. How insecure, discouraged, irritated, stressed, and annoyed were you? (1—Very Low, 7—Very High)\\\hline
}
\vspace{-.3cm}
\end{table*}

\begin{table*}
\begin{tabular}{l} 
\hline
Q1.1. Which robot was easier to communicate with? (1-Green, 7-Blue)\\
Q1.2. How did your communication strategy differ between the two robots? (Short textual answer)\\
Q2.1. Which robot was more mentally demanding to communicate? (1-Green, 7-Blue)\\
Q2.2. Which robot made you feel hurried or rushed during the task? (1-Green, 7-Blue)\\
Q2.3. Which robot made you feel successful in accomplishing the task? (1-Green, 7-Blue)\\
Q2.4. For which robot did you work harder to accomplish your level of performance? (1-Green, 7-Blue)\\
Q2.5. With which robot made you feel more insecure, discouraged, irritated, stressed, and annoyed? (1-Green, 7-Blue)\\\hline
\end{tabular}
\caption{After finishing all the tasks, participants comparatively rated (or a 7-point Likert scale) their communication experience (questions 1.1 - 1.4) and the subjective workload using NASA TLX (questions 2.1 - 2.5) comparing between the two robots. Participants rated on a 7-point scale. Green robot is pragmatic, and blue robot is non-pragmatic.}
\label{tab:poststudy}
\Description[Post-study survey questions]{Post-study survey questions:
Q1.1. Which robot was easier to communicate with? (1-Green, 7-Blue)\\
Q1.2. How did your communication strategy differ between the two robots? (Short textual answer)\\
Q2.1. Which robot was more mentally demanding to communicate? (1-Green, 7-Blue)\\
Q2.2. Which robot made you feel hurried or rushed during the task? (1-Green, 7-Blue)\\
Q2.3. Which robot made you feel successful in accomplishing the task? (1-Green, 7-Blue)\\
Q2.4. For which robot did you work harder to accomplish your level of performance? (1-Green, 7-Blue)\\
Q2.5. With which robot did you feel more insecure, discouraged, irritated, stressed, and annoyed? (1-Green, 7-Blue)\\
}
\vspace{-.3cm}
\end{table*}

\section{User Study Results}

%We present the quantitative and qualitative findings from the the user study that shed light on the research questions from the previous section.

\subsection{Quantitative Results} \label{sec:quants}

%We evaluated two design choices using our mixed method user study. The results are summarized visually in Figure \ref{fig:success_violin}, \ref{fig:examples_violin} 

\subsubsection{Task Completion Rates}

\begin{figure}[h]
  \centering
  \includegraphics[width=\linewidth]{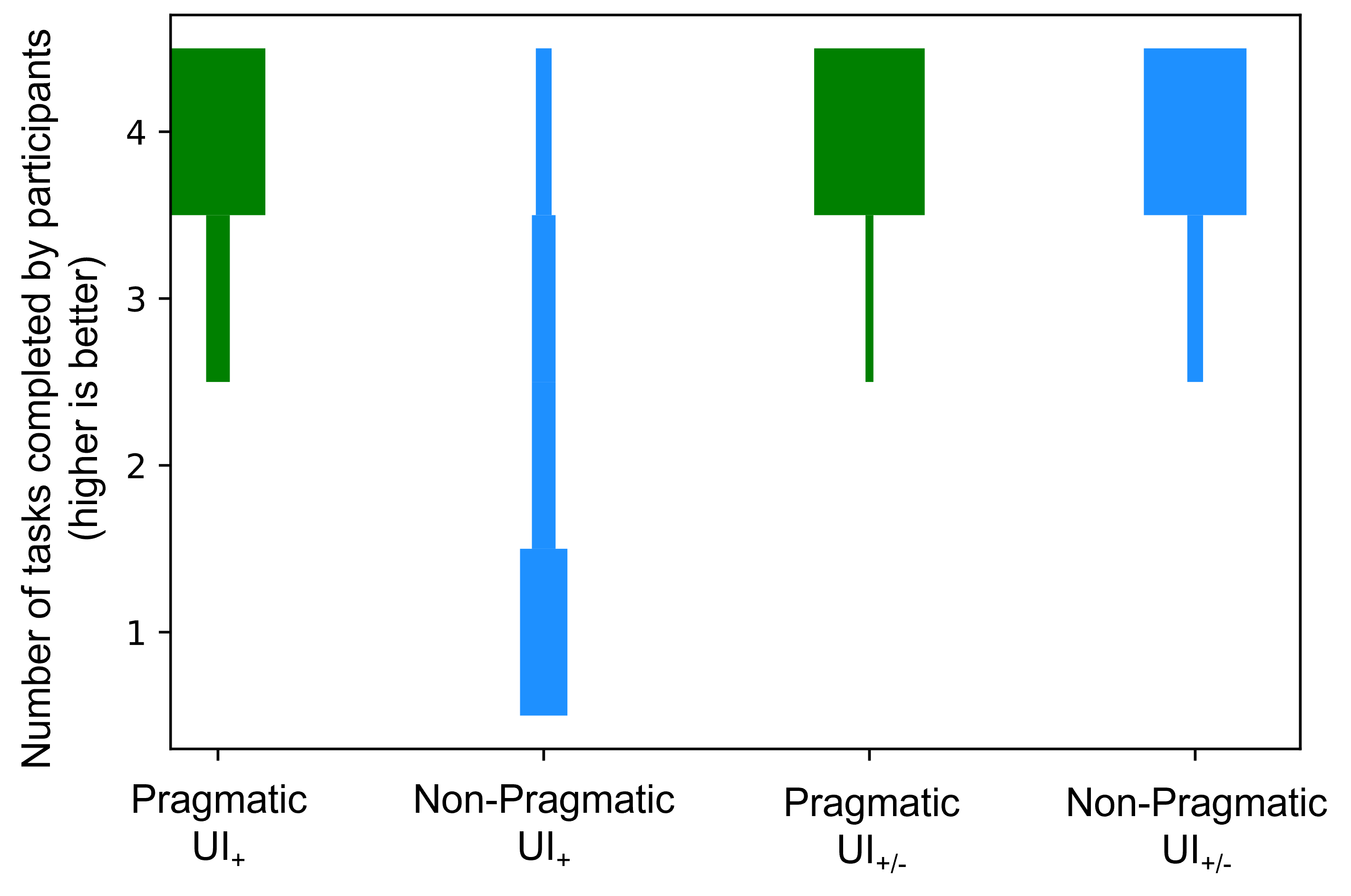}
  \caption{Violin plot of number of tasks completed in each condition. The pragmatic synthesizer performed well at guessing the intended regex regardless of interface affordances, while the non-pragmatic synthesizer needed both positive and negative examples to perform comparably well.}
  \label{fig:success_violin}
\end{figure}
% \todo{I'd change the word "task" (not the title, but the "2 out of 4 conditions" part) to something more like "condition" here, as task is 1 participants 20 minute kinda thing} hmm maybe i'm wrong here
Figure~\ref{fig:success_violin} shows that the pragmatic synthesizer was consistently good at guessing the right regex (a median of 4 out of 4 tasks completed), regardless of whether or not the participant was able to give negative examples in addition to positive examples. In contrast, the non-pragmatic synthesizer was very poor at guessing the right regex (a median of 2 out of 4 tasks completed) unless the participant was able to provide both positive and negative examples to clarify what they were trying to communicate, at which point participants reached a comparable rate of task completion (median of 4 out of 4 tasks completed) as with the pragmatic synthesizer. 
%More specifically, the task completion rates of participants communicating with a pragmatic synthesizer were very high (a median of 4 out of 4 tasks completed), and participants communicating with a non-pragmatic synthesizer . %to clarify what they were trying to communicate to the synthesizer. In contrast, 

%\todo{Figure 4 and 5 should be discussed in the order that they appear}

Looked at a different way, over the course of the user study, 30 participants each attempted to communicate 4 randomly assigned regular expressions to a pragmatic synthesizer, for a total of 120 task attempts. Only 4 ($3.3\%$) of those attempts were unsuccessful. The same 30 participants each attempted to communicate the same 4 randomly assigned regular expressions to the non-pragmatic synthesizer, and 33 of those attempts (over 8 times as many) were unsuccessful ($p < 0.001$).
The majority of these communication failures happened in the $\mathbf{UI_{+}}$ condition, where participants could only provide positive examples. In the same $\mathbf{UI_{+}}$ condition, participants successfully completed $95\%$ of communication tasks (57/60) to the pragmatic synthesizer, while only completing $48.33\%$ of communication tasks (29/60) with the non-pragmatic synthesizer ($p < 0.001$). In the $\mathbf{UI_{+/-}}$ condition, the difference between task completion rates between the synthesizers was not significant ($98.33\%$ for the pragmatic synthesizer vs $96.67\%$ for the non-pragmatic synthesizer), since the participants could give negative examples to refine the non-pragmatic robot's guess.

\subsubsection{Number of Examples Required}

\begin{figure}[h]
  \centering
  \includegraphics[width=\linewidth]{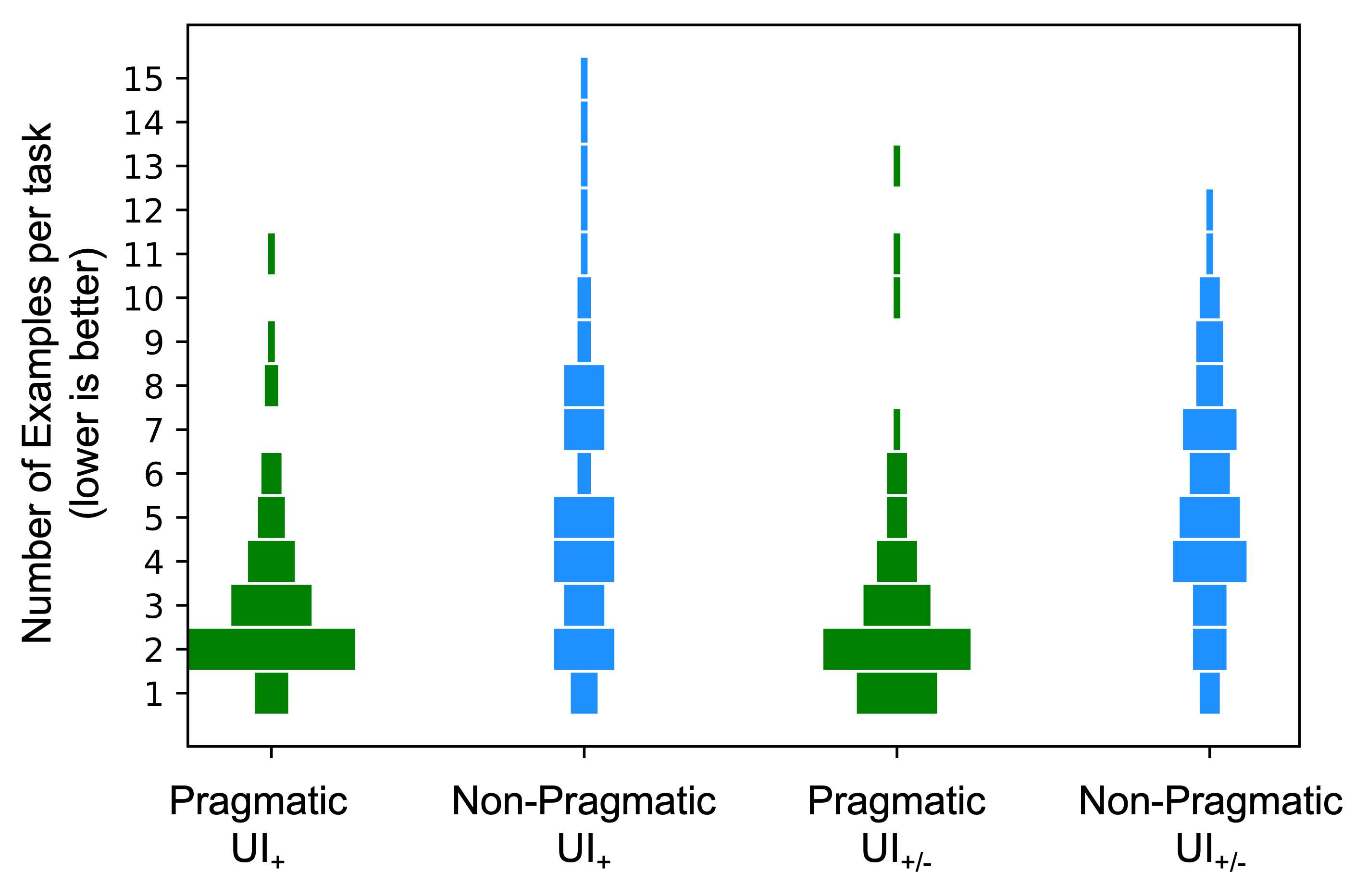}
  \caption{Violin plot of number of examples participants provided in each communication task. Regardless of interface affordances, participants provided significantly less examples when communicating with the pragmatic synthesizer compared to the non-pragmatic synthesizer.}
  \label{fig:examples_violin}
\end{figure}

Participants used significantly fewer examples ($p<0.001$) to communicate the regex successfully when communicating with a pragmatic synthesizer than when communicating with a non-pragmatic synthesizer. Specifically, participants communicating with a pragmatic synthesizer only required a median of 2 examples per task ($\mu=3.14$; $\sigma=0.22$) compared to participants communicating with the non-pragmatic synthesizer who required a median of 5 examples per task($\mu=5.46$; $\sigma=2.99$). The distribution over examples per task required in each condition is shown in Figure~\ref{fig:examples_violin}. This significant difference holds, regardless of whether the participant can provide negative examples or not. %This difference is statistically significant ($p<0.001$).

When comparing participants communicating with the same synthesizer but with different interface affordances, the ability to provide negative examples did not have a significant impact on the number of examples participants provided during the task.

%While the ability to provide negative examples had a significant impact of participants' overall task success rates, this affordance did not significantly change the number of examples participants provided during the task, on their way to success (or failure).

%If we look at the UI conditions individually, similar pattern follows. In $\mathbf{UI_{+}}$ condition, participants provided significantly fewer examples ($p<0.001$) with a median of 2.5 examples per task when communicating with the pragmatic synthesizer ($\mu=3.2$; $\sigma=2.02$) compared to a median of 5 examples per tasks when communicating with the non-pragmatic synthesizer ($\mu=5.4$; $\sigma=3.3$). Similarly, in the $\mathbf{UI_{+/-}}$ condition participants provided significantly fewer examples ($p<0.001$) with a median of 2 examples per task ($\mu=3.08$; $\sigma=2.4$) when communicating with the pragmatic synthesizer compared to a median of 5 examples per task ($\mu=5.5$; $\sigma=2.62$) when communicating with the non-pragmatic synthesizer

\subsubsection{User Behavior} \label{sssec:quant_behavior}

\begin{figure}
    \centering
    \includegraphics[width=1.0\linewidth]{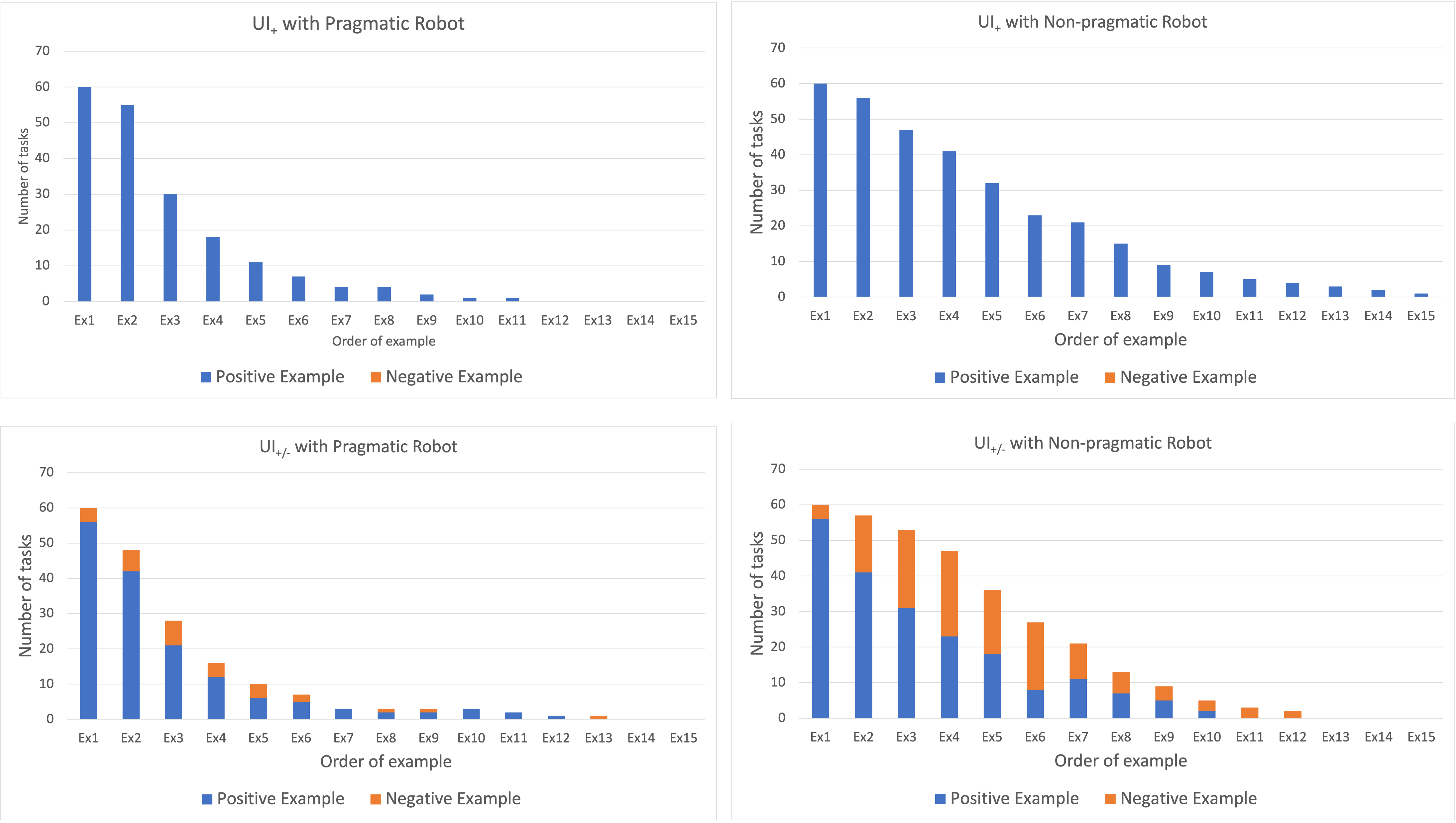}
    \caption{Number and proportion of positive and negative examples users provided until the synthesizer guessed the right regex.}
    \label{fig:ordinal}
\end{figure}

In the $\mathbf{UI_{+/-}}$ condition, where participants had the option to provide negative examples, they still provided more significantly more positive examples than negative examples on average (Fig. \ref{fig:ordinal}) by a factor of 2.33, when averaged across both synthesizer types ($p<0.001$). If we just consider the examples provided to the pragmatic synthesizer, this becomes more pronounced: participants provided 4.9 times more positive examples than negative examples. 

\begin{comment}
- X \% of participants started with positive ex with prag, whereas as Y for non-prag...
- X followed up with negative for prag, whearas ....
- for 50\% of tasks 
\end{comment}

Figure \ref{fig:ordinal} shows that ,in the $\mathbf{UI_{+/-}}$ condition, participants started with positive examples for 86.6\% of the tasks. However, for the following example, participants provided positive examples for 87.8\% of the tasks when using pragmatic synthesizer, and only  71.9\% for the non-pragmatic synthesizer. This difference becomes more pronounced if we consider the first three examples. Participants using the pragmatic synthesizer provided positive examples for the first three examples 80\% of the time in contrast to 48.3\% of the tasks for participants using the non-pragmatic robot. 

Given the fact that the pragmatic robot reasons about positive example as explained in Section~\ref{sec:pragsynth}, and the participants provided more positive examples to the pragmatic synthesizer, participants using pragmatic synthesizer finished 50\% of the total tasks with just 3 examples. Whereas, participants using the non-pragmatic synthesizer had to provide 6 examples to finish 50\% of the total tasks (Fig \ref{fig:ordinal}). In fact, participants communicating with non-pragmatic robot in the $\mathbf{UI_{+}}$ condition, where they could only provide positive examples, failed to complete 23.3\% of the tasks after they had exhausted all possible positive examples.

% Fig. \ref{fig:ordinal} also shows that the users generally started with a few positive examples, looked at the robot's guess and followed up with additional positive or negative counterexamples. In fact, in the $\mathbf{UI_{+/-}}$ condition, users started with a positive example for 86.6\% of the tasks and for 64\%  of the tasks the first three examples were positive (Fig. \ref{fig:ordinal}).\todo{rewrite this paragraph about the relationship between positive examples and pragmatic synthesis, focusing not on first positive examples (where folks have no info yet on the behavior of the synthesis algorithm) but on the follow-up examples being more positive in the pragmatic case} 

% Given how the pragmatic synthesizer reasons about positive examples,\todo{if this isn't already explained earlier in the section on pragmatic synthesis, add it and then add a reference to the section here and summarize the main point, i.e., that the synthesizer can get a lot out of positive examples} it is guaranteed to reach the target regular expressions with just positive examples.\todo{move this to the section about pragmatic synthesis, because it's not a result of your user study} In contrast, the non-pragmatic robot may not always reach the target regex with just positive examples, and it may require more negative examples. In fact, participants communicating with non-pragmatic robot in the $\mathbf{UI_{+/-}}$ condition, where they could only provide positive examples, failed to complete 23.3\% of the tasks after they had exhausted all possible positive examples.

\begin{figure}
    \centering
    \includegraphics[width=1.0\linewidth]{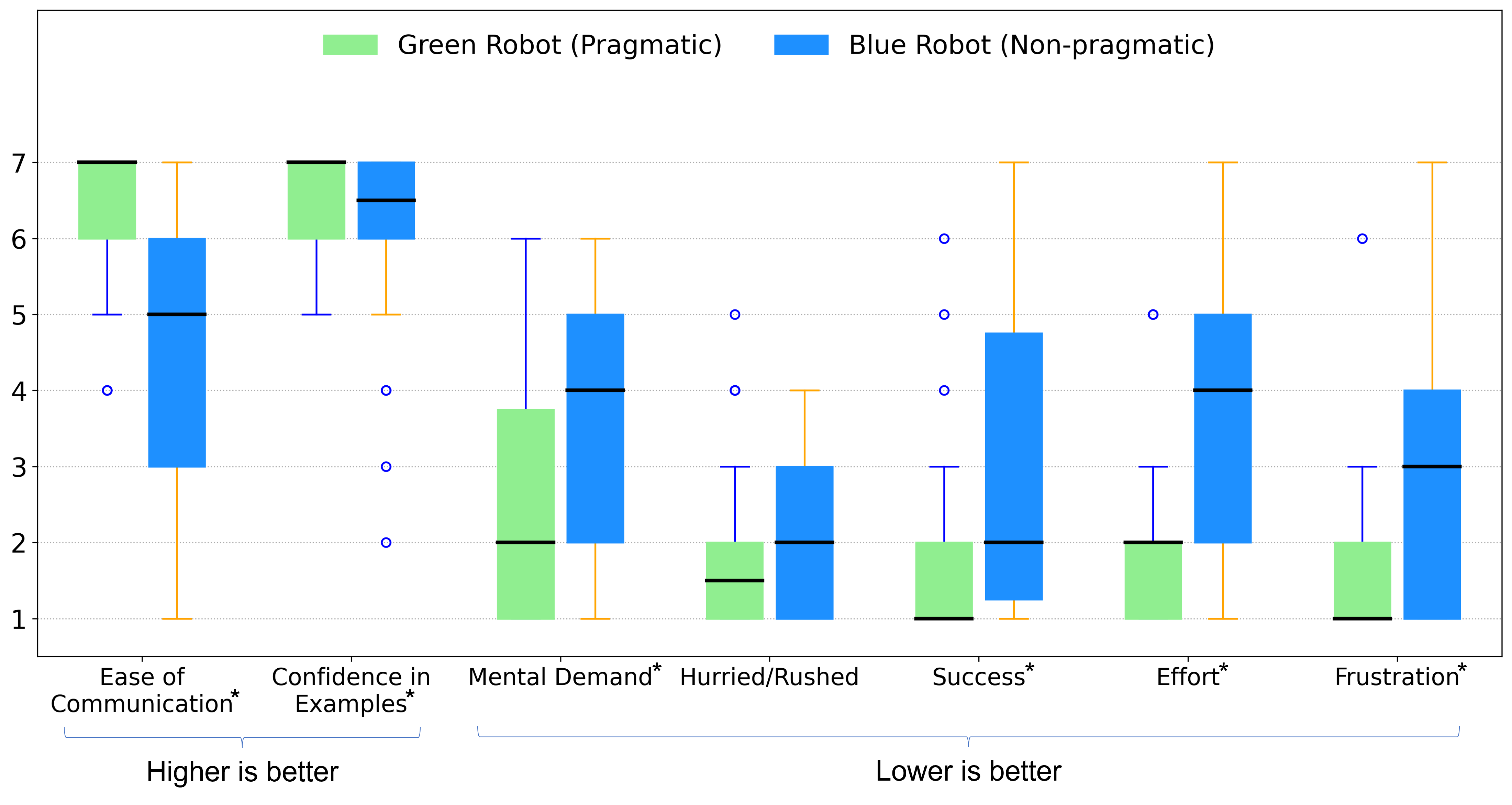}
    \caption{Self-reported scores measured in the post-task survey. (See Table~\ref{tab:posttask} for exact wording of questions. * represents differences that are statistically significant)}
    \label{fig:tlx}
\end{figure}

\begin{figure}
    \centering
    \includegraphics[width=0.6\linewidth]{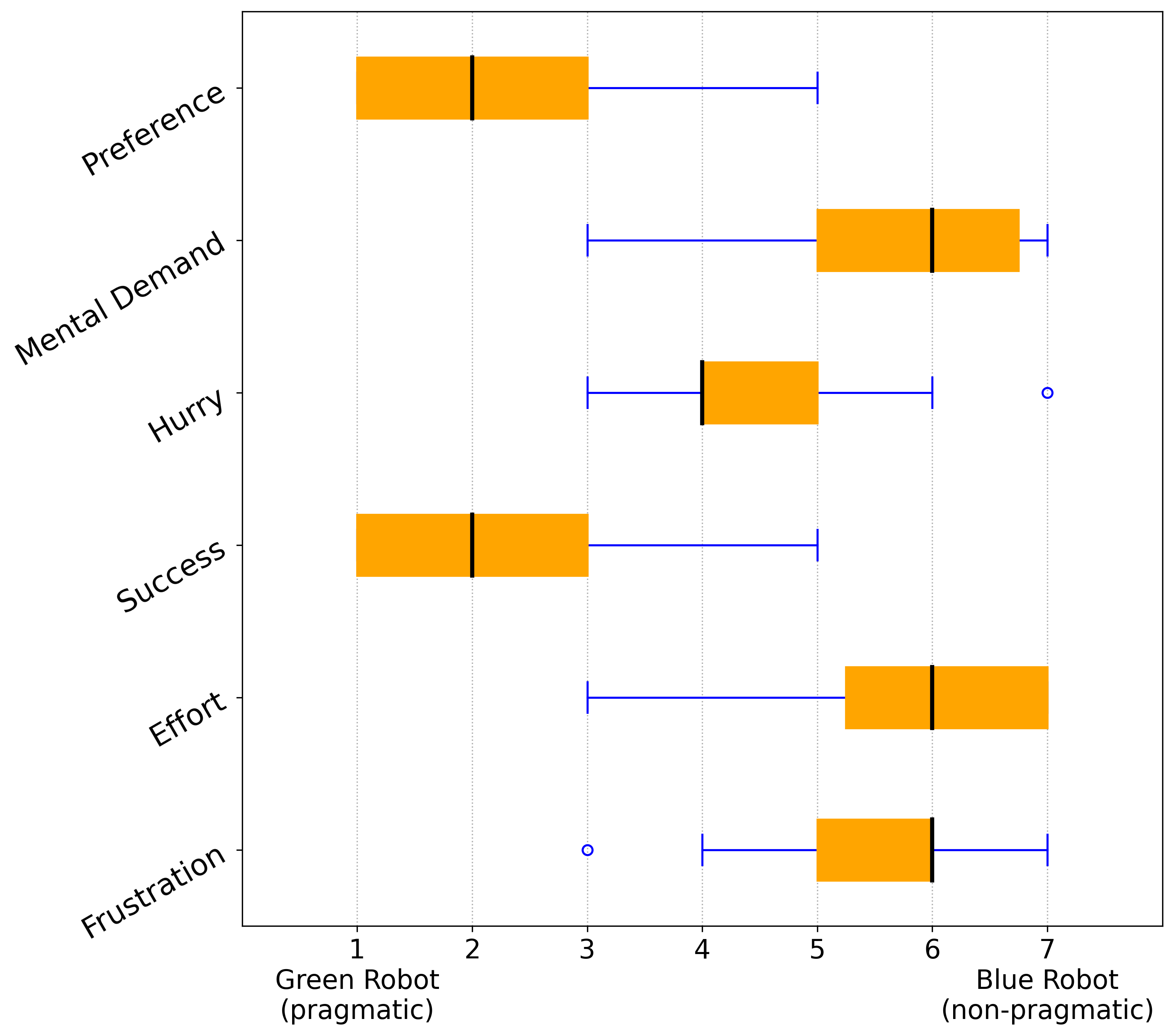}
    \caption{Self-reported scores measured in the post-study survey. (See Table~\ref{tab:poststudy} for exact wording of questions.)}
    \label{fig:poststudy}
\end{figure}

\subsubsection{Quantitative Subjective Ratings}
Figure~\ref{fig:tlx} shows the post-task quantitative ratings participants gave on the relative difference in quality of communication they experienced between the two types of synthesizers including five NASA TLX questions. When communicating with the pragmatic robot, participants felt significantly less mental demand ($p \approx 0.001$). felt less frustrated ($p < 0.001$), spent less effort ($p < 0.001$), and gave themselves better performance rating ($p \approx 0.002$). However, there is no significant mean difference in how hurried they felt while communication with the robots. Furthermore, participant using pragmatic robot found it significantly easier to communicate ($p < 0.001$, MANOVA), and felt more confident with the examples they provided ($p = 0.028$).

Figure~\ref{fig:poststudy} shows the post-study quantitative ratings participants gave  on the relative difference in quality of communication they experienced between the two types of synthesizers including five NASA TLX questions. When communicating with pragmatic robot, $83.3\%$ of participants felt less mental demand and less frustrated, $86.6\%$ of participants felt more successful and $86.6\%$ participants spent less effort compared to the non-pragmatic robot. There was no significant difference on how hurried participants felt when communicating with both the robots. $86.6\%$ of the participants reported they prefer the pragmatic robot over the non-pragmatic robot.

\subsection{Qualitative results}
We analyzed the participant responses from the post-task and post-study surveys and audio transcripts of when they thought aloud in order to get qualitative insights into the difference in experience of using a pragmatic or non-pragmatic synthesizer. 
 
\subsubsection{User behavior}

As described in section~\ref{sssec:quant_behavior}, users generally started with a few positive examples, and then gave positive examples or negative counterexamples based on the robot's guess (Fig.~\ref{fig:ordinal}). P11 said, \say{\textit{Starting with positive examples is very important. If the robot guess is very close, then I can give more negative examples to get to the precise regex.}} Some participants noticed that the pragmatic synthesizer could work off of positive examples extraordinarily well: \say{\textit{I used the strategy of starting with positive examples with both robots, but it seems that the [pragmatic] robot was able to guess well with only positive examples as compared to the [non-pragmatic] robot which required many positive and negative examples}} (P20).

Quantitatively we had shown that non-pragmatic robot required more negative examples to reach the target regular expression (Section~\ref{sssec:quant_behavior}). Participants highlighted this explicitly in the post-study survey responses. From the post study survey responses, 10 of 15 participants (P3, P7, P8, P11, P15, P19, P20, P24, P27, P30) in the $\mathbf{UI_{+/-}}$ condition felt they had to give more negative examples to eliminate incorrect guesses and constrain the search space to lead the non-pragmatic synthesizer to the target regex, whereas they believed the pragmatic synthesizer reached the target regex with few (or no) negative examples. P27 said,\say{\textit{The [pragmatic] robot would learn just enough from the positive examples to get to the regex. The non-pragmatic robot needed negative examples to narrow it down.}}

During the study, five participants (P2, P7, P23, P26, P27)---all of whom were interacting with the non-pragmatic synthesizer---blamed their own actions, not the robot's performance, for their inability to succeed in the tasks. \say{\emph{Am I doing something wrong?}} P2 thought aloud, when the non-pragmatic robot failed to guess the regex after multiple examples. Similarly P7 said \say{\emph{I think I'm not adding very good examples---it is getting worse}}. 
%All of the five instances of blame happened during their interaction with the non-pragmatic robot. 
Because in our study design, each participant had to separately communicate the same target regex to both the pragmatic and non-pragmatic synthesizers, we know that the five participants were not in fact bad at giving examples: all five successfully communicated the same target regex to the pragmatic robot.

%One possible explanation could be due to the inherent (high) expectation of good performance while performing the user studies. However participants blaming themselves even if it is due to the poor performance of the tool has been observed in prior work before \cite{huassuage}. 

\subsubsection{The value of negative examples}

%Participants wanted to provide negative examples. 
\say{\emph{I wish I could provide some negative examples}}, P2 thought aloud. In fact, six of 15 participants (P1, P2, P5, P13, P18, P26) in the $\mathbf{UI_{+}}$ condition wished they had the affordance to provide negative examples to help the robot reach the target solution during a failed task. Of these six participants, five were communicating with the non-pragmatic robot at the time, while only one (P18) was communicating with the pragmatic robot.

\subsubsection{Repeating examples}

Six participants (P4, P5, P6, P13, P21, P22) wanted to repeat an example they already provided during repeated failed attempts to get the non-pragmatic synthesizer to guess the right regex. P13 said \say{\emph{I want to give the same example again to see if it changes it's guess.}} When asked for their reasoning, P5 said they wanted to \textit{emphasize} an example more. P21 expected the robot to learn something new (gain more info) with the repeated examples after seeing all the other examples. P22 assumed the robot was using some probability distribution in the back-end, and it would increase its confidence in the example. P4, P13, and P22 wanted to repeat the example just out of spite since they had already exhausted all the examples they could provide. Participants wanted to give the same example multiple times in the hope of recovering from the failure.

\subsubsection{Mental demand of giving counter-examples.}
In the think-aloud study, P24 said \say{\textit{I had to think a lot harder for the blue robot [non-pragmatic synthesizer] than the green robot [pragmatic synthesizer]. For Green [the pragmatic synthesizer], I just had to think of the most clear-cut positive examples---which generally works. For blue robot[the non-pragmatic synthesizer], I had to think of a lot of negative examples.}} We can see that participants associated giving negative examples with working harder. One of the participants (P27) took it as a challenge saying \say{\textit{I liked coming up with negative examples for the blue [non-pragmatic] robot. Once you make it guess the regex, there was a sense of accomplishment}}. Furthermore, four participants (P4, P8, P9, and P12) blindly enumerated as many positive examples as they could without looking at the robot's guesses. They only start giving counter-examples after they had given many positive examples. These observations suggest that participants found it more intuitive and less mentally demanding to provide positive examples than negative examples. Prior research on concept learning has also shown that people can meaningfully generalize concepts on positive examples alone~\cite{xu2007word}. %Though we cannot conclusively say participants prefer giving positive examples over negative examples. Our results point towards an interesting research questions that requires further investigation. 

%% file: sec_discussion.tex
\section{Discussion and Future work}

%In this paper, we implemented pragmatic synthesizer for a small sub-domain of regular expressions to understand its usability. 
%\todo{write a topic sentence that summarizes or introduces the results discussed specifically in this paragraph}
The pragmatic synthesizer was, from both a user experience and performance perspective, superior to the non-pragmatic synthesizer in this regular expression sub-domain.
Participants found it easier to communicate with, regardless of whether the interface afforded just positive or both positive and negative examples, requiring fewer examples to communicate the intended regular expression.
%in both the conditions (with just positive examples - $UI_{+}$, and combination of positive and negative examples -$UI_{+/-}$) \textbf{\emph{[RQ1, RQ3]}}. 
%They also needed fewer examples to communicate successfully with the pragmatic synthesizer, compared to the non-pragmatic synthesizer.
%, they also successfully communicated with the pragmatic synthesizer using fewer examples \textbf{\emph{[RQ1]}}. 

Surprisingly, unlike the participants' experience with the non-pragmatic synthesizer, negative examples did not increase participants' success rates beyond what they could achieve with positive examples alone.  %participants were just as effective in their communication efforts when using only positive examples compared to giving both positive and negative examples. 
This is excellent from a user experience perspective because 
%This may be due to two reasons: (1) The recursive reasoning allows the pragmatic synthesizer to understand concepts with just positive examples (2) Users found it more natural to give positive examples \cite{xu2007word} to describe a concept. 
the quantitative and qualitative results indicate that users prefer giving positive examples and find it mentally more demanding to give negative examples. 
%We believe there could be more focused studies, both in regular expression domain and other programming domains, to understand the preference of users when it comes to providing positive versus negative examples. 
That said, 
%even though users provided significantly more positive examples compared to negative ones and, could effectively communicate with the pragmatic synthesizer with just positive examples, 
we still believe the ability to provide negative examples is important in order to clarify one's intent. Under certain circumstances, adding negative examples was more convenient for users to express their intent correctly to the computer. %Any user interface that supports synthesis---pragmatic or not---should allow the users to provide negative examples, even if the users choose to exercise it rarely.

There are also many limitations to pragmatics which we want to highlight. Currently, the formulation of synthesis as a communication task operates on a fixed number of concepts (individual regular expressions) and utterances (example strings), while traditional program synthesis often operates over a combinatorially large domain. In theory, we can scale up the number of concepts and utterances to millions of regular expressions and example strings, but the computational overhead currently limits its practicality. 
%Through this work, we better understand the potential draw attention to this pragmatic inductive bias worth investigating. 
Given the superior usability of pragmatic synthesis demonstrated by prior work~\cite{pu2020program} and our work, we believe (1) there is value in generalizing pragmatics-based inductive bias to additional domains to further test its general applicability, and (2) the computation methods necessary to scale pragmatic synthesis to larger domains are worth investing in.
%future work needs to address the scalability of pragmatic synthesis and its practical applications. 

Finally, to keep the comparison simple and focused to the research goals, we restricted the utterances to simple example strings. However, alternative forms of utterances such as semantic and data augmentation (\cite{zhang20,hila2018}) may also be compatible with future pragmatic synthesizers, further increasing the user's ability to quickly and effectively specify what program they want without writing it out. This would change the current formulation of pragmatics, but could be very helpful in synthesizing longer and more complicated regular expressions. In general, building pragmatic program synthesizers for the various form of utterances that are intuitive to end-users is an exciting, open research challenge.

%% file: sec_appendix.tex
%\subsection{An Incremental Pragmatic Synthesis Example}

%\begin{comment}
\section{The Incremental RSA Algorithm} \label{ap:algorithm}
%\todo{move it to appendix} 
\subsubsection{\textbf{Literal Listener $L_0$}} We start by building the literal listener $L_0$ from the meaning matrix [Table~\ref{tab:mm}]. Upon receiving a set of examples $D$, $L_0$ samples uniformly from the set of consistent concepts:

\begin{equation}
    \label{eq1}
    P_{L_0} \propto \mathds{1}(w \vdash D), \  P_{L_0}(w|D) = \frac{\mathds{1}(w \vdash D)}{\sum_{w'\in H} \mathds{1}(w' \vdash D)}
\end{equation}

applying to our example in [Table~\ref{tab:mm}], we see that $P_{L_0}(w_0|u_1, u_2) = \frac{1}{2}$ and also $P_{L_0}(w_3|u_1, u_2) = \frac{1}{2}$.

\subsubsection{\textbf{Incremental Pragmatic Speaker $S_1$}}

We now build a pragmatic speaker $S_1$ recursively from $L_0$. Here rather than treating $D$ as an unordered set, we view it as an ordered sequence of examples and models the speaker’s generation of D incrementally, similar to auto-regressive sequence generation in language modeling \cite{sundermeyer2012lstm}. Let $D = u^1  ... u^k$, then:

\begin{equation}
    \label{eq2}
    P_{S_1}(D|w) = P_{S_1}(u_1, ..., u_k|w) = P_S(u_1|1)P_S(u_2|w, u_1)...P_S(u_k|w, u_1, ... u_{k-1})
\end{equation}

where incremental probability $P_S(u_i|w, u_1, ..., u_{i-1})$ is defined recursively with $L_0$: 

\begin{equation}
    \label{eq3}
    P_S(u_i|w, u_1, ..., u_{i-1}) \propto P_L(w|u_{1...i}), \  P_S(u_i|w, u_1, ..., u_{i-1}) = \frac{P_{L_0}(w|u_1, ... u_i)}{\sum_{u_i'}P_{L_0}(w|u_1, ..., u_i')}
\end{equation}

Applying this reasoning to our example in Table~\ref{tab:mm}, we see that $P_{S_1}(u_1, u_2|w_0)$ is:

\begin{equation}
    \label{eq4}
    P_S(u_1|w_0)P_S(u_2|w_0, u_1) = \frac{P_{L_0}(w_0|u_1)}{\sum_{u'}P_{L_0}(w_0|u')} \frac{P_{L_0}(w_0|u_1, u_2)}{\sum_{u''}P_{L_0}(w_0|u_1, u'')} = \frac{4}{11} \mbox{*} \frac{3}{7} = 0.156
\end{equation}

\subsubsection{\textbf{Pragmatic Listener $L_1$}} Finally, we construct an pragmatic listener $L_1$ which recursively reasons about the pragmatic speaker $S_1$:

\begin{equation}
    \label{eq5}
    P_{L_1}(w|D) \propto P_{S_1}(D|w), P_{L_1}(w|D) = \frac{P_{S_1}(D|w)}{\sum_{h'}P_{S_1}(D|w')}
\end{equation}

In our example, $P_{L_1}(w_0|u_1, u_2) \approx 0.66$, $P_{L_1}(w_1|u_1, u_2) = 0$, $P_{L_1}(w_2|u_1, u_2) = 0$, $P_{L_1}(w_3|u_1, u_2) \approx 0.34$. As we can see, the intended concept $w_0$ is ranked the highest, in contrast to the uninformed listener $L_0$.
%\end{comment}